This paper is a preprint of a manuscript submitted to EuroVis 2025 that was not accepted for publication

# Interactive Analysis of Global Explanations using Aggregated Class Activation Maps for Network Data

I. Cherepanov[1], D. Sessler[1], A. Ulmer [1], F. Wagner[1], T. May [1], J. Kohlhammer[1,2]

[1]Fraunhofer IGD, Germany
[2]TU Darmstadt, Germany

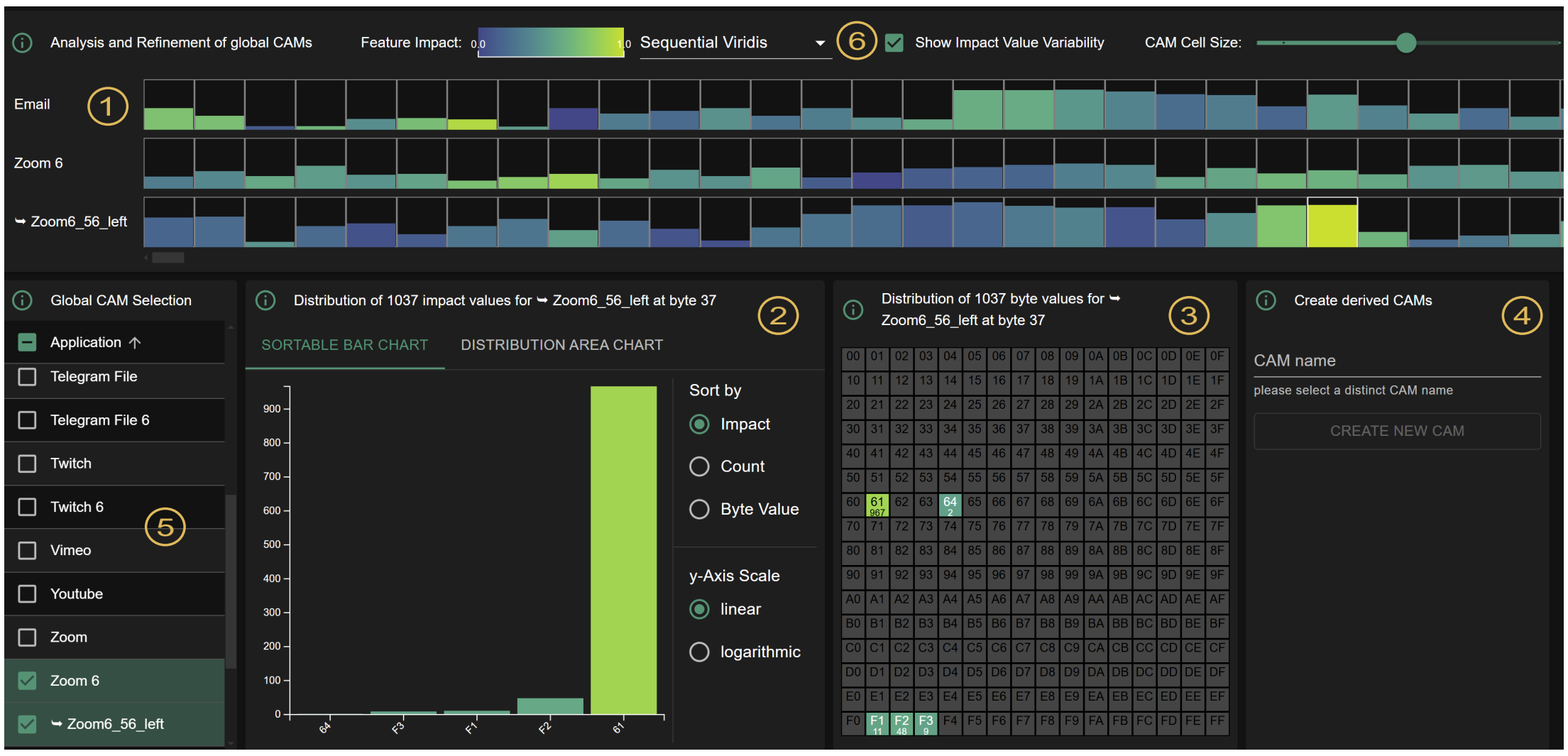

Figure 1: Overview of global explanations (1) for different classified application classes (5). A global class activation map (CAM) has two indicators: color and size. The color indicates the influence of each byte, while the bar size reflects the variability across samples. The interactive visualizations, which present different settings (6), assist experts in identifying the most significant data for deeper analysis. Experts can refine a global CAM by exploring various visual representations that highlight the impact values of a specific application (2-4).

**Abstract**

*Recent machine learning (ML) advances have demonstrated that deep learning (DL) achieves impressive results in different application domains, including the classification of computer network traffic to corresponding applications. However, the data frequently contains diverging patterns within a single predicted class. This presents a significant challenge to the ability to provide a clear and comprehensive explanation and emphasizes the necessity for tools capable of detecting and analyzing these patterns. Furthermore, the capacity to extract descriptive rules for classes is a crucial requirement in network traffic analysis and intrusion detection, particularly when leveraging advanced tools like next-generation firewalls. We provide a visual-interactive system that explains predictions of classes for network traffic. Global explanations derived from multiple samples of a given class contribute to understanding model predictions. Visualization of global explanations enables recognition of different patterns that offer experts a more comprehensive overview of its characteristics. We introduce a prototype that facilitates visual exploration and refinement of global explanations, enabling network experts to detect and refine new patterns for specific applications. These explanations support the identification of misleading features and the formulation of new rules for the management of networks. Our approach also aims at enabling ML experts to acquire new insights, including the possibility of separating or merging classes and the development of more accurate and reliable DL models. Our proposed prototype was evaluated by experts in machine learning and network analysis.*

## 1. Introduction

The accuracy achieved through the use of machine learning (ML) techniques in various fields is remarkable, surpassing human accuracy in numerous tasks. Hence, artificial intelligence (AI) methods are being integrated into complex systems to support experts in achieving their goals [DKAK20]. In particular, when data complexity or high dimensionality impede the ability of humans to discern patterns or relationships. At the same time, there is a growing demand for understanding model decisions, not only in safety-critical solutions, but also in fields such as finance or network analysis [NVK*15, RvGH18, BGG*21]. Understanding AI models is crucial for ensuring transparency, accountability, and trust in their decisions. It helps to uncover biases, identify potential errors, and enables experts to interpret and validate predictions effectively, thereby facilitating knowledge extraction and decision making [RBAW*23, SJB*22].

Explainable artificial intelligence (XAI) approaches play a pivotal role in enhancing the interpretability and explainability of ML methodologies. There exists a challenge of superior-performing ML models being more complex, thus lacking in explainability [AB18, DDN*23]. Additionally, an overarching challenge lies in aligning ML and XAI methodologies with human objectives, principles, values, and cognitive processes [YCY*21, WWH*23, WZTWQ22]. To tackle this obstacle, visual analytics (VA) utilizes visualization methods to convey information to humans and facilitate abstract perception and deduction [Cui19]. Consequently, it becomes feasible to communicate model decisions to domain experts using XAI. This bridges the gap between various domain experts and ML specialists, fostering knowledge exchange, and enhancing the accuracy and reliability of AI applications.

Our paper addresses this challenge in the field of Network Traffic Monitoring and Analysis (NTMA) by focusing on the classification of traffic data [AST21] and visualizing the contribution of class-specific features [HDK*21, AS22]. This approach empowers domain experts to create rules based on payload information for advanced systems such as Suricata, Snort, and Next-Generation Firewalls (NGFWs) [Fou24,Sys24]. The rapid growth and variability of network data necessitate accurate traffic classification for advanced network management tasks such as QoS provisioning, pricing, mitigating cyber threats, and supporting regulatory requirements.

Deep learning (DL) outperforms traditional statistical methods by autonomously learning hidden patterns in network traffic, eliminating the need for manual feature extraction. Recently the architecture of a one-dimensional convolutional neural network (CNN) performed best in network classification [LJSSHZS20]. The advantage of this architecture is that an explanation can be calculated directly during the classification of a single sample, known as a class activation map (CAM) [ZKL*16]. The CAMs consist of rectangles, each representing a byte, with the color indicating the byte's influence on the classification [CUJK]. The advantage of the semantically structured nature of network data, where the position of each unit of information remains constant, lies in its ability to aggregate local explanations for features across all samples [CSU*23]. This aggregation represents a global explanation of a class.

Overall, this paper contributes: (1) A visualization of an XAI technique with local CAMs, aggregated and presented after data classification by the model. These global explanations enable network experts to identify patterns and key points of interest within a class, while facilitating comparative analyses across different classes. This process allows the expert to gain new insights into the data and to support the extraction of descriptive rules through global explanation. Furthermore, the model's capacity to illustrate its own decision-making processes enhances comprehension, fostering increased trust in the system. (2) We also provide the user with a visual interactive approach to refine the aggregations by eliminating samples. This functionality is crucial because numerous patterns within a class may overlap or cancel each other out. The visualization of the impact value distribution allows the user to detect different patterns in a class. The advantage of an intuitive visualization of explanations is that it facilitates the exchange of information between experts from different backgrounds [JSS*18, EAJK*]. (3) Our proposed XAI visualization system was evaluated by two groups of network and ML experts.

## 2. Related Work

Our review of the relevant literature is organized into two parts. In Section 2.1, we delve into the field of network traffic analysis and explore techniques for this specific data type. Following this in Section 2.2, we investigate work centered on the classification of network data, culminating in a summary of XAI techniques within this field.

### 2.1. Network Traffic Analysis

Analyzing network traffic is crucial for enhancing network security [DDM*19, LSF*15, MKN*07]. It assists in identifying potential threats such as unauthorized data flow and allows for the formulation of effective prevention strategies. A crucial component of network analysis is the emphasis on identifying network traffic. This area concentrates on the capability to precisely analyze the different types of traffic in a network. Established tools such as Wireshark [CC10, MMD*] empower network experts by allowing them to filter network data through command-line interfaces, focusing exclusively on data relevant to their diagnostics. Visualization techniques enable users to explore networks more effectively through visual representations of data and its characteristics [MKN*07, JJJ23, JCC*23]. Interactive visualizations, such as statistical tools, charts, and diagrams, are widely used to assist domain experts in understanding network traffic data. VA has been demonstrated to be an effective approach for tasks such as anomaly detection in network traffic [CvW] and identifying patterns within network packet flows [CMEvW, USK21]. Ulmer et al. [USK] introduce a visual interactive tool tailored for network experts, which supports data drill-down via filtering of packet header properties, such as IP, port, and time, which fosters engagement with network data. While these tools facilitate thorough data analysis, they fall short in packet classification tasks due to the unreliability of traditional methods that solely rely on ports or IP addresses for packet characterization. Additionally, certain applications might use dynamic port allocation and masking, using port numbers assigned to trusted applications. Due to these challenges, advanced techniques are needed to accurately classify contemporary network traffic.

### 2.2. Classification of Network Traffic with XAI

Payload inspection or deep packet inspection (DPI) analyzes packet headers and payloads for application classification using predefined patterns and regular expressions tied to certain applications [EMAEBE, AN13]. However, maintaining high prediction accuracy is challenging due to the rapid evolution of applications, requiring frequent updates to patterns and difficulty recognizing new ones. Deep learning (DL) addresses this by autonomously learning features, eliminating the need for manual rule extraction [LBH15, CKS*09]. Subsequently, a variety of approaches were investigated, employing a range of DL architectures. The work by Lotfollahi et al. [LJSSHZS20] conducted a comparative analysis between a stacked autoencoder (SAE) and a 1D CNN, named Deep Packet, for the classification of encrypted traffic. CNN architecture contains convolutional hidden layers featuring a series of independent filters (kernels) responsible for convolution operations. This approach based on CNN architecture outperformed previous approaches, which was confirmed by Montieri et al. [ACMP19] who tested various models in their study. Based on this, we also utilize a DL model of this architecture in our work.

DL models, due to their complexity and the vast number of parameters involved, have one disadvantage: a lack of interpretability [Rai20]. The use of such models is constrained by the absence of trust resulting from the lack of explainability of the classifications, which is especially true in critical fields like cyber security [SJB*22]. To tackle this challenge, a technique called class activation maps (CAM) was introduced, which is based on CNN [ZKL*16]. This explanation technique originated in computer vision and aims to explain the decisions made by a CNN model. A CAM represents a saliency map highlighting the regions of an image that contributed the most to the prediction of a particular class. The resulting CAM is calculated directly from the trained model, meaning that no approximation model is required, making the process computationally cheap. In our previous work [CUJK], we adapted the CAM-based XAI technique to classify network packets and interpret individual classifications. These CAMs were then visually represented as a 2D color-coded grid representing the bytes alongside the original packet in hexadecimal representations. The positions in the CAM were linked to the bytes of the packet. This allowed the network analyst to select the impacting bytes on the CAM and observe their linkage to the segment in the packet, as well as to the part of the hexadecimal representation. We later demonstrated that CAMs can be aggregated when applied to semantically structured data [CSU*23]. Semantically structured data has a fixed format and order, as seen in network data with predefined packet structures. Aggregating local CAMs provides a global explanation of classes. In this work, we apply this approach to the classified network data, utilizing the aggregated local CAMs. In contrast to the visualizations of the local CAMs in previous works, we now illustrate the global CAM in a line as a one-dimensional sequence of color-coded bars. This visualization is inspired by scarf plots [YW] which facilitate the effective comparison of structurally similar data in a compact visual form. This enables an efficient explanation of differences between distinct classes in long packet data. We propose a visual-interactive solution to refine CAMs, aiming to uncover distinct patterns within a class and support the creation of descriptive rules for these patterns.

## 3. Data-User-Task

Within this section, we initially describe the format of PCAP files. Subsequently, we introduce the targeted user groups, along with their respective tasks in PCAP analysis. This serves as the foundation for deriving the requirements for our approach.

### 3.1. Data

Packet capture (PCAP) files store captured network traffic packets in a binary file format. They are generated by network monitoring tools such as Wireshark or tcpdump [CC10, Tcp09]. PCAPs enable network administrators, security analysts, and researchers to examine network traffic patterns, identify anomalies, troubleshoot network issues, and detect potential security threats. Each packet within a PCAP file includes a header containing metadata such as the timestamp of the packet, the length of the captured packet, and other relevant information about the network communication. Following the header, the packet payload is stored in its raw binary format, encompassing various network layer protocols of which each has its own header again. The headers contain information such as the source and destination addresses, protocol version, packet length, time-to-live, checksum, and other control information necessary for the transmission and routing of the packet across the network. The last payload in each packet, also known as the packet's data section, contains the information transmitted over the network. This includes data from high-level application protocols, such as website content, email messages, file transfers, or any other type of data exchanged between devices on the network. The structure of network data is often compared to an onion with a header for each layer, followed by the rest as payload. This nested design results in semantically structured data, ensuring that the position of each piece of information remains consistent [CSU*23].

### 3.2. Users

Within the scope of our approach, we prioritize the involvement of network admins and analysts dedicated to comprehensive network management and examination on an expert level. Our proposed technique aims to identify patterns for the formulation of new rules for advanced next-generation firewall systems. This user group also includes malware analysts and cybersecurity professionals in general because they can benefit from our approach by acquiring knowledge and understanding the characteristics of application classes.

The second user group we target is ML developers, whose task is to provide models for network analysts. This group benefits from our approach by exploring the visualized explanation of the classifications, making model decisions more understandable. Finally, our system provides a foundation for the communication between both user groups, facilitating improvements to the AI system. In this context, the expert users of our approach are:

- **Network experts:** Administrators, Cybersecurity professionals, Malware analysts
- **Machine learning experts:** ML developers, AI researcher

### 3.3. Tasks

Network experts rely on a variety of tools [CC10, MMD*] to analyze captured network data for securing infrastructure, monitoring behavior, detecting threats, and optimizing network performance. Identifying patterns and defining descriptive rules are crucial for these tasks. Platforms such as MITRE ATT&CK and *ThreatFox.abuse.ch* provide pre-defined rules based on payload information, enhancing threat detection capabilities. These rules are applicable in advanced systems like Suricata, Snort, and Next-Generation Firewalls (NGFWs), which leverage them for more effective threat identification and mitigation.

Network data can also be captured and labeled into categories, forming the foundation for ML models, which excel in classifying incoming traffic. However, these models often act as black boxes, limiting their trustworthiness and interpretability despite their strong predictive performance.

To address these challenges, we propose leveraging XAI techniques to extract actionable insights from ML models. Our approach aims to generate rules directly from captured data, bridging the gap between automated model predictions and human-interpretable security protocols. Based on our experience in this area, examination of related work [BEK14, FB14, EFW*, XML*18], and exchange with domain experts, we take all these aspects into account and address the following tasks of the two target groups. The first two tasks are of network analysts, the next two tasks are of ML experts, and the final task involves collaboration between both groups:

**T1:** Optimize Quality of Service (QoS) and manage network resources.
**T2:** Develop and enforce security policies and procedures to ensure compliance with industry regulations and standards, including regulating the inbound and outbound flows of applications.
**T3:** Provide AI models and techniques that explain model decisions, enabling understanding and acquisition of potential knowledge.
**T4:** Identify and address biases or misleading features to ensure the correctness of the ML model.
**T5:** Assist in optimizing AI models by identifying areas for improvement.

### 3.4. Design Requirements

Based on the user tasks we derived several requirements for our approach:

**R1:** Display an overview of class explanations (**T1-T3**).
**R2:** Identify the most influential bytes in each class (**T1-T3**).
**R3:** Comparison of byte-wise sequences within each class (**T2, T3**).
**R4:** Visualization of the impact value distribution for a specific byte (**T4, T5**).
**R5:** Refinement of a class explanation to detect different patterns (**T4, T5**).

## 4. Visual-interactive XAI System

In this section, we present the technical aspects of our AI model, encompassing model architecture, training and the classification process. Following this, we proceed to explain the local predictions, describing the methodology behind the calculation of these explanations. Subsequently, we aggregate these local explanations to calculate the global explanations of classes. This includes a discussion on the rationale behind the design of our visual analytics system, the methodology used to compute a global CAM, and the principles guiding such aggregation. Finally, we discuss the visual representation of these explanations and the interactive components provided to enhance user engagement and insight discovery.

### 4.1. Network Traffic Classification XAI System

In this work, we utilize a DL model based on a 1D-CNN architecture, whereby 1D refers to the dimensionality of the input data being processed. To address this, we utilize 1D kernels in our CNN model to capture relationships between neighboring data points along the same axis effectively. There exist precisely two compelling rationales for employing this ML architecture in the analysis of network data. The first one is that this model performs best on the network data [LJSSHZS20, ACMP19], the second reason is the possibility of calculating local explanations (local CAMs) during classification [CUJK]. The local CAM calculation is based directly on the trained model, making it a computationally cheap method. It enables us to derive explanations straight from the model itself, avoiding the need for approximations that are commonly required by other explanation techniques, like SHAP or LIME [RSG16, LL17].

The use of DL techniques for network packets requires additional preprocessing steps. Network data is variable due to its nested structure of different protocols, consequentially the packets differ in size accordingly. In neural network architectures, the input layer mandates fixed-size input, thus demanding preprocessing measures to homogenize network packet sizes by truncating the data of longer packets and padding the shorter ones, similar to the work of Lotfollahi et al. [LJSSHZS20]. The reason for limiting the payload to 1500 bytes is that the maximum transmission unit (MTU) typically restricts the payload in computer networks to this size. Ethernet headers provide important details about physical connections, like MAC addresses. However, they do not give information about higher-level applications, and they are not used for classification. Finally, to prevent the model from making inaccurate decisions based on IP addresses, both the source and destination IP addresses within packets are uniformly set to zero.

We trained a 1D-CNN model with the following feature map dimensions in the hidden layers: 16, 32, 5 layers with 64, and 128, utilizing 1500 inputs in all layers, with a stride size of 1 and a kernel size of 7. Once we had a good baseline of this model on the balanced ISCX VPN-nonVPN benchmark dataset [fC15], we trained it on our own balanced dataset with new application classes. Our model attained an F1-score of 97%.

The CAM technique aims to identify regions within an image sample that significantly contribute to the predicted class. It works by extracting feature maps from the final convolutional layer of a CNN model. Once these feature maps are obtained, a method called global average pooling (GAP) is applied. GAP calculates the average value of each feature map, which is then passed to the final fully-connected layer. By mapping the weights of the output layer onto the convolutional feature maps, the resulting CAM is computed as the sum of all convolutional feature maps from the last convolutional layer, each multiplied by the corresponding weights of the output layer. This process is illustrated in Figure 2. CAMs highlight the regions within the input data that have the strongest influence on the final classification decision. This process enables visualization of the areas of interest in the input data that contribute most significantly to the model's prediction. In our case, the input data is one-dimensional, representing bytes. The CAM technique remains applicable, with the added distinction that the kernels used are now one-dimensional. At the end of the process, we obtain the CAMs, which identify the specific bytes within a packet that have the greatest influence on the final classification decision.

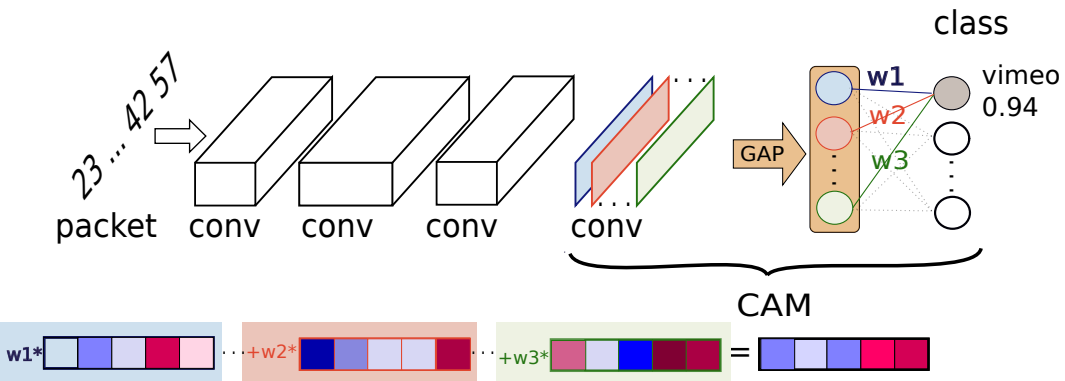

Figure 2: A CAM is calculated by mapping the predicted class score back to the last convolutional layer of a CNN model. This is achieved by multiplying the convolutional feature maps of the last layer by the weights of the output layer and then summing them up

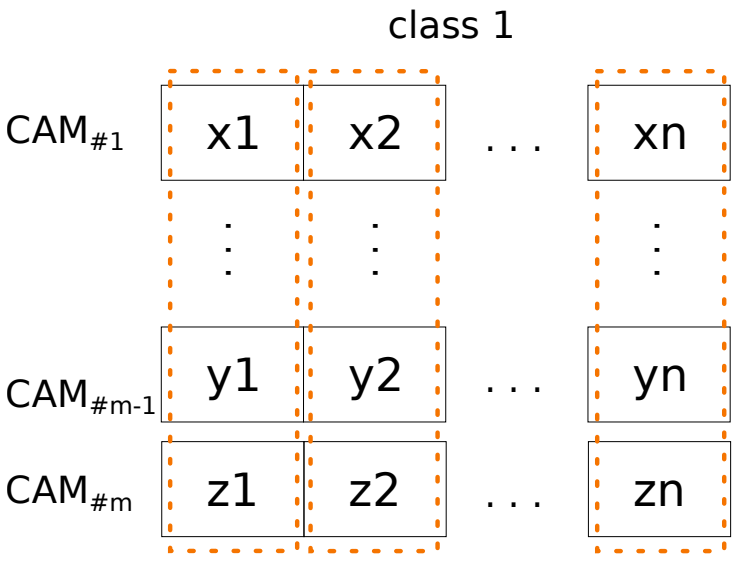

Figure 3: Illustration of global CAM calculation: The aggregation of values along the vertical axis from multiple local CAMs, from which the two indicators are calculated (color and size). $n$ represents the number of features in a local CAM, and $m$ represents the total number of local CAMs involved.

A CAM explanation provides insight into the classification of individual samples, making it a local explanation. The next step involves aggregating these local explanations to construct a global explanation for a specific class. The rationale behind considering aggregations is grounded in the semantic structure of the data. Specifically, each data attribute consistently maintains a fixed position and length within a packet of the same type. To create the global CAM (Figure 3), we aggregate the values along the vertical axis, condensing the importance of each feature into a single value. Various aggregation methods are provided to capture different aspects of the impact distribution in classification analysis. One such method is computing the mean, which provides an overall average of the distribution. This approach was considered appropriate for our case because it represents the overall average of the distribution of impact values (R2). Alternatively, the median of the CAM values serves a similar purpose but is less influenced by extreme values, which may act as outliers. Additionally, we compute a secondary indicator to quantify the variability of feature impacts across all CAMs for the individual class (R2, R4). This indicator assesses whether the range of impact values predominates significantly across all local CAMs of the class or if the values are rather dispersed. Between the tested variability measures, entropy was most suitable, because it measures the purity or randomness of the impact values among local CAMs. The final step involves normalization within a class. All normalization settings must remain consistent, even when a CAM is generated based on a subset of samples from this class, thereby ensuring uniformity across analyses.

### 4.2. Visualization of Aggregated CAMs

We employ visualizations to present a comprehensive view of classes with 1500 features each (1500 bytes), enabling users to identify interesting patterns easily (R2). For the representation of distributions, box plots or violin plots are often utilized. However, although these plots are effective for displaying and contrasting various distributions of values, they are limited to comparing a relatively small number of distributions. Given our requirement (R4) to compare a hundred or more distributions, we sought an alternative approach. To enhance visualization scalability, we transform distributions into two key indicators, represented in each grid cell by a colored bar: aggregated value and variability (R2). In the work by Andrienko et al. [AA23] and in our previous approach [CSU*23], similar aggregations are represented as a square patch. Square patches are used in data visualization, but they have disadvantages compared to bar charts. The main limitation of square patches is that they rely on area comparison, which is less perceptually accurate than length comparison. Bar charts are better for tasks requiring precise or relative value interpretation [CM84, War12, Mun14]. We assign the aggregated impact values to a sequential colormap, facilitating analysts' ability to quickly spot patterns or areas within the lengthy vector that significantly influence the class prediction. We opted for a sequential colormap as illustrated in Figure 1(1), where bytes with the highest aggregated impact values are colored yellow, those with the lowest are colored blue. This color scheme effectively marks the bytes with the greatest and least influence on classification (R2). We considered various color schemes commonly used including jet, viridis, and turbo [LH, RS21]. After conducting a comparison, we decided not to use the jet scheme due to its inadequate brightness gradient and its lack of accessibility for individuals with color blindness. We decided to provide multiple colormaps as a user parameter including viridis (as default colormap) and turbo for their broad color spectrum and white-blue and white-red colormaps for their simplicity of interpretation. The impact variability value is represented by the size of the bar within each grid cell (R1).

In contrast to our previous work [CUJK], where the CAM for individual predictions was visualized, this approach presents the global CAM in a novel manner (Figure1(1), 5). The visualization of individual predictions arranges bytes linearly from left to right, akin to the structure of text in a book, where each new line begins after a certain number of bytes. This method is particularly effective for detailed analysis of individual packets, allowing for a comprehensive examination of byte-to-hex code mappings displayed alongside. In contrast, our focus is on visualizing global class explanations, offering the ability to refine these and compare them with other classes directly (R3). Therefore, we chose to present the global CAM as a single vertical vector, displayed in one line. This representation method avoids the identification of semantically incorrect patterns by acknowledging that byte sequences are related horizontally rather than vertically (Figure 1(1)). Additionally, by juxtaposing multiple lines of global CAM representations we facilitate an efficient byte-by-byte comparison of different classes (R3).

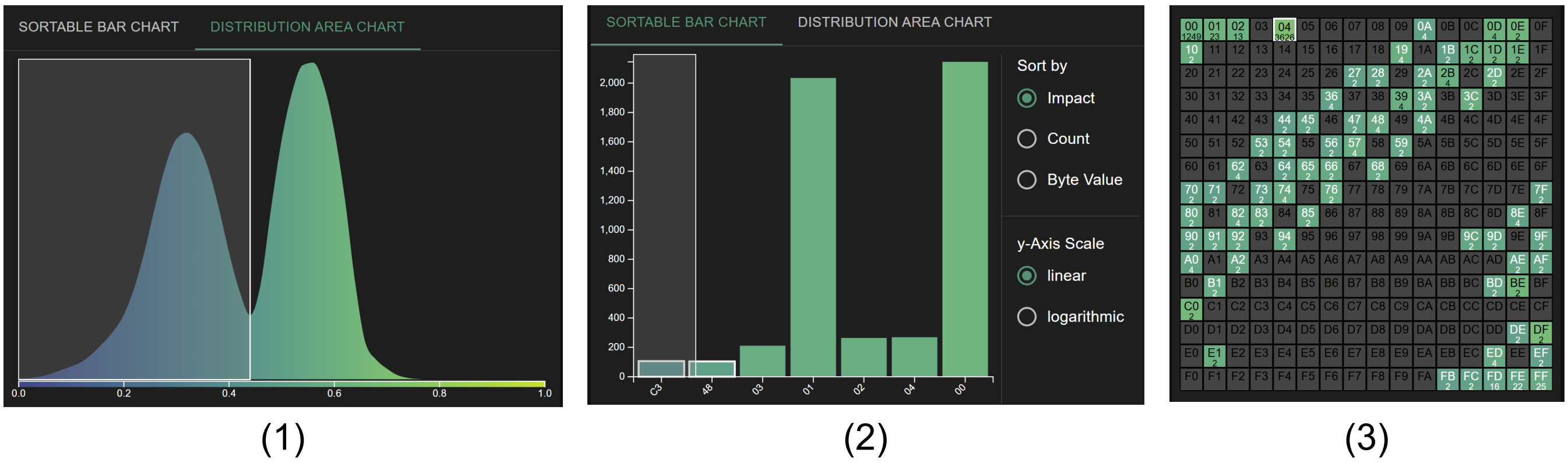

Figure 4: These visualizations that are independent, illustrating different aspects. The distribution of impact values is visualized (1). The byte values are illustrated on the x-axis, with their frequency on the y-axis, and the bars are colored according to their corresponding impact values (2). All possible byte values are shown, with the occurring ones highlighted and their impact values indicated by color (3).

### 4.3. Interaction and Refinement

In the interface, classes are listed on the left side, enabling users to select or deselect them according to their analysis needs, shown in Figure 1(5). This capability ensures that analysts can tailor their view to focus on classes of interest, streamlining the analysis process and enabling more focused investigations (R1). Additionally, analysts can utilize tooltips within the CAM, accessible by hovering the mouse pointer over one of the cells. Here, they can find the indexes of the cells and the indicators as numbers (aggregated value and variability), along with the number of samples used to calculate this CAM. A legend in the upper panel provides information about colors and sizes (Figure 1(6)). Horizontal scrolling is provided to adjust the visualization's centering on the horizontal axis to enable users to access all bytes in the sequence.

Analysts can adjust the size of the cells and access two levels of detail via a slider, shown in Figure 1(6). When the cells are larger than 10 pixels, both visually encoded attributes, aggregated impact, and variability, are enabled. For cell sizes from 10 pixels down to a single pixel, the variability attribute is disabled, and the cells turn into stripes with the selected width, while maintaining a fixed height to ensure visibility, as shown in Figure 5. In this manner, the visualization transforms into a scarf plot, a pixel-based visualization that scales even better for the visual analysis of thousands of values simultaneously. Adjusting the cell sizes within the CAM visualization offers several advantages for analysts seeking to interpret complex data (R1). Firstly, altering the cell sizes can provide a better overall view of the CAM, making it easier for experts to grasp the entirety of the data landscape at a glance. This enhanced visibility is particularly useful in quickly identifying the most influential bytes within a class (R1, R3). By varying cell sizes, analysts can compare full CAMs side-by-side which are aligned on top of each other (R3). This facilitates a clearer understanding of how different classes relate to each other.

Following an initial overview of the impact values, analysts have the opportunity to delve into specific byte for further exploration. Our methodology employs two key indicators to highlight such bytes of interest. Bars highlighted in yellow indicate a high degree of influence. However, shorter bar sizes for these bytes suggest a wider distribution of impact values (Figure 4(1)). Similarly, shorter bar sizes with a color not corresponding to the minimum or maximum values color are particularly noteworthy as they may represent bytes with multiple peaks in the impact value distribution. This approach facilitates a targeted exploration, enabling analysts to uncover and investigate bytes with unique distribution patterns in an efficient manner (R4, R5).

We implemented three different visualizations that illustrate the distribution of the impact and byte values, shown in Figure 4. If an expert identifies a byte especially interesting and decides it requires additional examination, a density plot of impact values is provided upon clicking the corresponding cell (Figure 4(1)). This visualization enables a detailed examination of the feature's impact distribution (R4). This density plot outlines all the impact values associated with this byte. A key benefit of this plot is its inclusion of color coding for the impact values (Figure 4). This capability enables the analyst to quickly discern which segments of the distribution are pivotal to the byte's influence and which values are comparatively irrelevant. In addition to the density plot, we provide a bar chart that visualizes the frequency of byte values, shown in Figure 4(2).

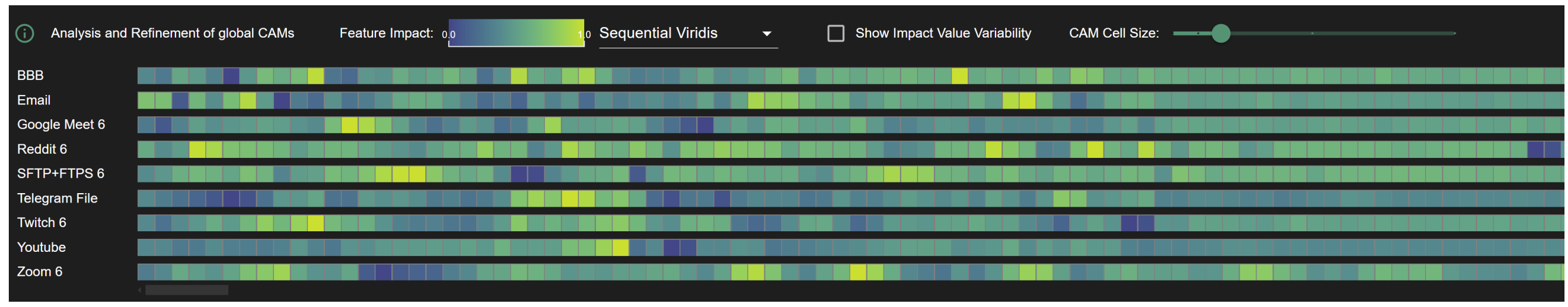

Figure 5: Adjustable cell sizes in our CAM visualization: This function allows users to modify cell size, enabling a comprehensive view of multiple CAMs simultaneously. For selected CAMs, enlarging cell sizes aids in a detailed examination of our second visualized indicator, the variability of impact values for a certain byte.

This chart allows for flexible exploration through various ordering options, including sorting by impact, count, or byte value. Furthermore, we offer logarithmic scaling to accommodate datasets with a wide range of impact values, enhancing interpretability and comparison. On the right side of the histogram, we display all possible byte values in a grid to illustrate the variability of values across the packets (Figure 4(3)). This visualization employs the same color-coding scheme to represent the impact of each byte value, we also indicate its frequency of packets through numerical annotations. This approach facilitates an intuitive understanding of how individual byte values influence the dataset while providing insights into their distribution and prevalence.

All three distribution visualizations facilitate filtering which complements each other. The histogram and bar chart using an area brush and the byte value grid by cell selection and deselection of distinct byte values (R5). Filtering actions initiate the computation of a new global CAM that incorporates the values selected from the specified class. Subsequently, the newly generated global CAM, which is named by the expert, is added to the class list. It is then displayed next to the existing global CAMs, enabling a comparative analysis. This functionality not only facilitates a deeper exploration of the data but also enables experts to isolate and analyze the impact of specific subsets of impact values. Additionally, it allows for the detection and extraction of varying patterns within a class. This refinement process can be iteratively applied across various bytes and even extended to newly created global CAMs, theoretically continuing until a single local CAM in the targeted class remains (R5). This iterative approach enables a step-by-step drill-down analysis that concentrates on specific aspects of the data and model behavior. When calculating a refined global CAM based on a subset of samples from a class, normalization is performed using the values from the entire set of samples within that class. This configuration is crucial for maintaining consistency and enabling meaningful comparisons between the refined global CAM and the initial comprehensive one. This approach ensures that even when analyzing a subset, the analysis remains aligned with the context of the broader class data.

## 5. Usage Scenario

We present two scenarios demonstrating the use of our visual-interactive system, aiming to assist network analysts in their investigative tasks. Additionally, it supports collaboration with ML experts to enhance model reliability through comprehensive global explanations of network traffic. While our primary focus is on the global explainability of classes, our tool also provides analysts with a visual representation of the data. It supports data filtering and integrates a classification model for categorizing the data [USK, CUJK].

Let us look at a common task for a network analyst. In the role as a **network analyst**, we often rely on Wireshark as the industry standard for analyzing captured network traffic [CC10]. However, with large datasets and intricate patterns, Wireshark can become slow and resource-intensive. Its manual filtering, interpretation, and analysis processes are time-consuming, limiting efficiency in handling complex network tasks. We prefer to use tools that provide a visual representation of the data and support filtering, allowing us to narrow down the data more quickly and make it more understandable [USK, CUJK]. Prefiltering options such as narrowing down to specific time ranges, selecting certain protocols, or eliminating broadcast and other irrelevant network packets are often crucial steps for effective analysis. Especially when detecting patterns within the same class, it is important to ensure that the data is consistent in terms of protocols, such as UDP and TCP, or, when extractable, higher-level protocols should be considered.

After uploading network traffic data to the proposed tool, we see a global explanation of the data, which highlights crucial bytes and their influence. When we see a long yellow bar in the visual representation, it strongly indicates that the corresponding byte is impactful for a particular class. These bytes become the primary focus of our subsequent analysis. Analyzing packets based on these impactful bytes facilitates the development of a characteristic specification for the class. Next, we turn our attention to less prominently highlighted bars, as these may be candidates for refinement. If a class exhibits multiple clusters of impact values that cancel each other out during aggregation, we use the drill-down method provided by the tool. We recognize cells containing such disturbances by observing the length of the bars, which reflects the variability of values in the aggregation. Shorter bars indicate a strong variability of the impact values between the local CAMs. This can occur, e.g., if a byte of a class is significant in many samples, but in others this byte is not relevant for the classification. For example, this could be information in the header that is not present in all network packets of a class. In this case, we iteratively refine a global CAM for a specific class by interactively focusing on a cell of interest with a smaller bar. By selecting the cell, we can observe the distribution of local CAM values. If the histogram reveals multiple modes in the distribution of CAM values of a specific byte, it is useful to separate them from each other, as illustrated in Figure 4(1). Then we filter the histogram by selecting a sigle mode. Choosing the single mode ensures a similar distribution of impact values. Consequently, we isolate packets with distinct impact characteristics, as shown in Figure 4(1). Through this process, we are able to uncover new information and identify novel patterns specific to the class, thereby delineating new characteristics or even defining a specific subclass within the original class. This iterative approach enhances the depth and precision of the network analysis. After this analysis, we can create rules for IDS like Suricata, based on the identified patterns and impact values. To illustrate, we are looking for unique values for a byte, as shown in Figure 1(3), while Figure 4(3) shows an ambiguous byte with different values within it. For simplicity, an example of a constructed rule is illustrated below, where specific byte indices are mapped to particular values.

```
alert tcp any any -> any any (msg:"APP3";
    content:"|3D|"; offset:37; depth:1;
    content:"|01|"; offset:62; depth:1;
    content:"|14|"; offset:70; depth:1;
    sid:1000004; rev:1;)
```

Listing 1: Example rule for Suricata IDS

This rule is configured to analyze TCP traffic with both source and destination IP addresses and ports set to 'any', meaning it applies to all network traffic. When triggered, the rule generates an alert with the message *APP3*, signaling the detection of a defined pattern. The simplified rule looks for three specific byte sequences within the packet payload. The first sequence, represented as |3D|, must be located at byte offset 37, with the search limited to a depth of one byte. The second sequence, |01|, is searched from byte offset 62, also limited to one byte. The third sequence, |14|, is checked at byte offset 70, with a similar depth restriction. Such tools are designed to inspect data at higher-order TCP/IP communication layers, including the application layer, enhancing their capability to safeguard network integrity.

During the analysis of significant bytes, we may encounter situations where we recognize inconsistencies or semantic errors made by the model. For example, IP addresses are often masked or the Ethernet header is completely excluded from the classification. This tool allows us to identify features that could be misleading or with which we disagree. This recognition is crucial for refining the model's accuracy and ensuring its reliability. If we discover such semantic discrepancies, we can communicate these findings with the ML specialist. This communication is facilitated by our ability to directly address the problematic feature. We can identify such byte numbers and mark which ones are misleading for a class or general, which ones should be masked, and which ones should be excluded completely. Working with AI tools that empower us as a network analyst also enhances the capabilities of ML specialists by meeting specific model requirements, enriched by our domain knowledge. Furthermore, the process of establishing a more reliable AI system step by step simultaneously grows our trust in these systems.

Let us look at a common task for ML professionals. We have limited knowledge of network data analysis, so we rely on the exchange of information. We can leverage the tool to enhance our understanding of how ML models interpret network traffic data and gain insights into the model's behavior, strengths, and limitations. By examining the global explanations generated by the tool, we gain insights into the features and attributes that are most influential in the model's predictions. This understanding of classification and the exchange of information with network experts helps us refine and optimize our models, improving their accuracy and performance. For instance, empirical evidence from prior studies indicates the necessity of excluding the Ethernet header and IP address from classification to prevent erroneous class mapping [LJSSHZS20]. Other features that could be misleading could be eliminated after negotiation with network experts.

Additionally, we can use the tool to identify areas where the model may be biased or prone to errors, allowing us to implement corrective measures and ensure the model's reliability [HR17, Zha07]. The explanation provided by the tool gives us the flexibility to experiment with model parameters and analyze the model's behavior. For example, we conduct an analysis by initially training a model exclusively on two classes and subsequently training a model that includes additional classes. By following these steps, we can make a comparison to identify features with an impact and discern the differences in impact between the two models, focusing specifically on the initially chosen two classes. This process enables us to gain a deeper understanding of the model's dynamics and how its performance and interpretability change with varying class configurations. This approach is particularly beneficial for classes that are difficult to separate. Training a model exclusively on two classes offers more insights, as its performance is not influenced by samples from other classes. This method allows us not only to measure performance but also to assess the quality of explanations generated by the model. Such analysis can reveal how model complexity affects its ability to distinguish between challenging classes and the clarity of the explanations it provides, thereby informing optimization strategies for both accuracy and interpretability.

## 6. Evaluation

To assess the usability and effectiveness of our visualization system, we conducted an evaluation with two groups of experts from different domains. The methodology was informed by established practices in usability testing and visualization research, including task-based evaluation and the think-aloud protocol [AM18, LBI*12].

### 6.1. Evaluation Setup

The first group consisted of three experts proficient in ML. This group included two male and one female participants aged between 31 and 34 years, with professional experience ranging from 4 to 6 years in the ML field. The second group consisted of three professionals specializing in network analysis and cybersecurity. These participants, all male, were aged between 25 and 50 years, with expertise ranging from 2 to 20 years. Following this, participants engaged in task-based interaction with the system, where they were assigned tasks tailored to their respective areas of expertise. For the ML experts, the tasks focused on interpreting and analyzing model behavior through our visualization, while the network analysts concentrated on identifying impactful bytes and uncovering patterns within cybersecurity datasets [Shn, LBI*12]. However, in order to assess the core functionalities of the system, including the intuitiveness of our visual coding and interaction design, we included similar questions for both groups, phrased using terminology tailored to their respective domains. This approach ensured that feedback on fundamental aspects of usability could be collected consistently.

Each evaluation session followed a structured process. Participants were first introduced to the visualization system through a brief, 15-minute training session that provided an overview of the system and interaction mechanisms. This introductory phase ensured that all participants had sufficient understanding to independently explore the system during subsequent tasks.

During the task execution phase, we employed the think-aloud protocol to capture participants' thought processes as they interacted with the system. This technique allowed participants to verbalize their reasoning, intuitive interactions, and any challenges they encountered, providing detailed qualitative insights into the system's usability [Nie94]. Our questionnaire included Likert-scale questions with responses ranging from 1 (strongly disagree) to 5 (strongly agree), to assess specific aspects of the system such as usability, feature intuitiveness, and task efficiency [JKCP15]. Some questions also offered multiple-choice options or space for open-ended written feedback to capture nuanced user perspectives. All participants completed the System Usability Scale (SUS) questionnaire at the end of their session [Bro13].

The findings from the evaluation are presented in two parts. The first focuses on the network analysis and cybersecurity experts, detailing the domain-specific challenges and feedback identified during their sessions. The second part addresses the results from the ML experts, emphasizing their unique perspectives and the system's performance in supporting their tasks. This division ensures that the evaluation reflects the distinct requirements and expectations of the two user groups.

### 6.2. Expert Interviews

In both expert groups, we asked related questions to assess their understanding of our visualization mapping and its ability to identify impactful bytes. Specifically, we evaluated whether the experts could locate the bytes with the most influence in the packet. A general question inquired about the location of the impactful bytes, as well as whether the experts could recognize that the encrypted section did not reveal influential bytes. Additionally, we asked the experts to identify the most important byte within a class by providing its index number. All experts completed tasks well without much effort, identifying the most impactful byte. Next, the experts were tasked with selecting three classes and determining which two exhibited the most similar pattern in their sequence. All responses were correct, further indicating the clarity of our visual mapping. When asked whether specifying a rule for a specific class would be helpful, all experts agreed. We also assessed their understanding of the variability representation of impact values across bytes. Experts analyzed the distribution of impact values in the global CAM and were asked to identify which byte had the most influence. Again, all responses were accurate. They confirmed their understanding of the variability enconding and noted that the histograms were helpful in examining value distribution.

**Network Experts**: We asked if an expert says a byte is not important, but the model thinks it is, can our explanation help talk to an ML specialist. The response from all was yes, with the argument that the index references byte, aiding in the argument about what exactly is wrong. When asked if they had previously used a system capable of packet classification, the answer was no. They had seen a solution from Cisco that could classify data [BYLPR10]. However, this solution didn't provide any explanation about model decisions. We inquired whether packet classification providing classification explanation is meaningful. The response indicated strong agreement (5,5,4). On the question of whether network data classification in general is meaningful, showed a tendency towards strong agreement (4,5,5). A cybersecurity analyst immediately identified a use case related to one of the recent leaks, in which data had flowed from the company's network to the outside. If certain applications had been classified and not allowed to exit the company's network, the incident might have been prevented [mit]. The response regarding the level of trust in automatic classification generally fell in the middle range, indicating a tendency towards distrust (2,3,3). We also asked whether the purpose of the global CAM was understandable; the responses were positive, confirming its comprehensibility (4,4,5). In addition, the experts were asked to briefly describe what the purpose is, their descriptions were not quite precise ML-wise but met the goal of global CAM. We asked the experts if the global CAM made them trust our AI system more. Here, we received a divided opinion (3,4,5). One of the experts emphasized the importance of explanations, stating that they even see it as more beneficial for the model to acknowledge when it's wrong about something rather than just providing a classification. The question "Does the global CAM visualization help gain insights into how the classification works?" elicited mixed responses from the experts, with one not seeing added value in terms of insights (2,4,5). With two identical responses of (4,5,5) to the questions "Does the global CAM help validate the model?" and "Does the global CAM help improve the model?", this confirms the experts' expectation to validate and enhance the model using the technique. The experts found our visualization to be suitable for network packets.

**ML Experts**: The task for the ML experts was to put themselves in a scenario where they would receive labeled data from network experts and train a model on it. They would be provided with our XAI technique as assistance. We asked the experts what they would do if they discovered a classification error between two classes (Class 1 being classified as Class 2 and vice versa), and additionally found that the global CAMs of these classes were also similar. Some suggestions were made, such as the possibility of merging these classes in consultation with the network expert, increasing the number of examples of these classes during training, or training the model specifically on these two classes and then examining the global CAMs to see if they remain similar. We asked the experts to look at a byte with a bimodal distribution and queried them about the influence of this byte. All experts stated that because there are two peaks, one in the green area and the other in the yellow, this byte indicates a feature that is important for some samples and unimportant for others in the classification. The experts were asked to examine global CAMs of multiple classes in their entirety. It was evident to all that the rear part of the payload exhibited no patterns and was represented in a less significant color. When asked if they would exclude this part, responses such as "Without the domain expert, I wouldn't want to exclude anything" were received. Subsequently, the question arose whether this visualization would aid in communication with the domain experts. The responses (4,5,5) indicated agreement. When asked if they understood the purpose of the global CAM visualization, the responses were (4,4,5), confirming that the global CAMs are understandable. "The global CAM visualization helps to gain insights into the model's classification" received responses of (4,4,5), showing agreement leaning towards 5. The experts responded with agreement to the question of whether the model can be improved through the global CAM. However, in response to whether the model can be validated with it, they rated it (3,4,4), which indicates light doubts.

In both questionnaires, we included open fields asking the experts what they liked and what they would improve. What the experts liked was the intuitive usability and understandable settings on the dashboard. They also liked that the important bytes immediately caught their attention. The overview and interaction with the elements were also praised. Suggestions for improvement included offering a choice of different color scales. The network experts also mentioned that the initial learning curve is somewhat challenging for them, and they would need more information on all provided functionalities. This is mainly due to the many ML terms and background functionalities with which they are not familiar.

In evaluating the usability of our system, we used the SUS to quantify the user experience provided by our tool. The SUS gave us a score of 80.16 (ML experts) and 80.83 (network experts). The SUS questionnaire again showed that a good introduction to the topic was important for the experts, as the question about the need for support was confirmed by many.

## 7. Discussion, Limitations and Future Work

The evaluation showed that network experts have not been previously engaged with the application of models and explanatory techniques for classification, encountering a significant amount of novel information about ML techniques. For this reason, deeper and more detailed guidance is necessary for them. This group recognized the value of such applications, and the evaluation and the SUS score received positive ratings.
When such techniques should be employed in companies, specifying a desired ML model presents a challenge because each company focuses on different applications. Therefore, network experts must be involved in the creation of the training datasets. Furthermore, adding any new application classes requires a retraining of the model, during which the balance between classes should be carefully maintained, and data drift must also be considered as it can affect the model's performance.
In future, we aim to provide functionalities that allow experts to mask certain features (such as setting IP addresses to 0.0.0.0) that may be misleading. The changes must be integrated and correctly visualized when the model is retrained. Furthermore, our plan involves adding automatic suggestions of impacting bytes to the analyst. These suggestions will satisfy one of the criteria, such as identifying the most impacting bytes, those with second strongly separated peaks in distribution, and bytes that are completely unimportant. We will also analyze approaches that can automatically generate rules, by using the classification and its explanation [CBG*20, CDK*21]. We also aim to automatically analyze the distributions, enabling the rapid detection of outliers or imbalances among particular bytes. Additionally, sorting functions such as sorting by columns or rows are also considered. Since the CAMs can also be viewed as sequences, we plan to apply sequence mining techniques on the explanations [ME10].

## 8. Conclusion

In this paper, we introduced a visual-interactive system designed to serve as an interface between experts, algorithms, and data. We aggregated local explanations from a CNN model for network data into global explanations to interpret application classes. Additionally, we provided a summary of the background and related work in network analysis, classification of network data, and interpretable DL methods. We defined user groups and their tasks, deriving requirements for our system. Subsequently, we described our approach based on these requirements, introduced our visual-interactive explanation of the underlying AI system, and demonstrated it in two usage scenarios. Furthermore, we prepared questions and tasks on real datasets and interviewed experts, demonstrating that our tool is intuitive to use and experts recognize its added value. Through these interviews, we gathered valuable feedback that we plan to address in the ongoing development. Reflection and discussion on our work have led to new ideas that we aim to tackle in the future.